\documentclass[
english, aps, prd,
superscriptaddress,
preprintnumbers,
nofootinbib,
twocolumn,
10pt
]{revtex4-2}

\usepackage[utf8]{inputenc}

\usepackage{amssymb}
\usepackage{amsmath}
\usepackage{amsfonts}
\usepackage{graphicx}
\usepackage{siunitx}
\DeclareSIUnit\barn{b}

\usepackage{babel}
\usepackage{setspace}

\usepackage[colorlinks,linktocpage=true]{hyperref}
\hypersetup{citecolor=blue,linkcolor=blue,urlcolor=blue}

\newcommand{\be}{\begin{equation}}
\newcommand{\ee}{\end{equation}}

\newcommand{\commonCaptionA}{
    The number in parenthesis indicates the number of retained terms in the Legendre series, which is determined from the estimated Monte-Carlo uncertainties according to the truncation prescription described in Sec.~\ref{sec:truncation}.
}
\newcommand{\commonCaptionB}[1]{
    Distribution of #1 of Higgs boson produced in WBF, reconstructed using histograms (left) or Legendre moments (right). The blue, orange, and black lines correspond to the leading-, next-to-leading, and next-to-next-to-leading order result. The bottom panels show relative deviations from the leading-order one.
    For Legendre moments the number in brackets indicates the number of moments retained after the truncation procedure described in Section~\ref{sec:truncation}.
    The uncertainty bands account for Monte-Carlo integration uncertainty, and, for Legendre moments, the truncation uncertainty.
}

\allowdisplaybreaks

\begin{document}

\preprint{TTP26-001, P3H-26-002}

\title{On the reconstruction of kinematic distributions computed with 
Monte Carlo methods using orthogonal 
basis functions}

\def\KIT{Institute for Theoretical Particle Physics,
	Karlsruhe Institute of Technology, 76128 Karlsruhe, Germany}
\def\IAP{Institute for Astroparticle Physics, Karlsruhe Institute of Technology, 76344 Eggenstein-Leopoldshafen, Germany}

\author{Kirill~Melnikov}
\email{kirill.melnikov@kit.edu}
\affiliation{\KIT}

\author{Ivan~Novikov}
\email{ivan.novikov@kit.edu}
\affiliation{\KIT}

\author{Ivan Pedron}
\email{ivan.pedron@kit.edu}
\affiliation{\KIT}
\affiliation{\IAP}

\begin{abstract}
\noindent
Reconstruction  of one-dimensional kinematic distributions 
from calculations based on high-dimensional 
Monte-Carlo integration is a standard problem in high-energy physics.
Traditionally, this is  done by collecting randomly-generated events in histograms.
In this article, we explore an alternative approach, whose main  idea is to approximate  the target distribution
by a weighted sum of orthogonal basis functions whose 
\emph{coefficients}
are calculated using the Monte-Carlo integration.
This method has the advantage of directly yielding smooth approximations to  target distributions. Furthermore,  in the context of high-order perturbative calculations with  local subtractions, 
it eliminates the so-called bin-to-bin fluctuations, which often severely affect the quality of 
conventional histograms.
We also demonstrate that the availability of a high-quality  approximation to the target distribution, for example the leading-order result in the perturbative expansion, can be exploited to construct an optimized orthonormal basis.
We compare the performance of this method to conventional histograms in both toy-model and real Monte-Carlo settings, applying it to  Higgs boson production in weak boson fusion as an example.

\end{abstract}

\maketitle
\section{Introduction}
\label{sec:introduction}
Monte-Carlo methods play a key role in  high-energy physics \cite{James:1980yn, Lepage:1977sw, ParticleDataGroup:2024cfk}. A particular application of such methods that will concern us in this paper, is  the computation of   distributions of observables  in collider processes in the context of perturbative QCD \cite{Buckley:2011ms}. 

A key step in the usage  of these methods is to organize  the simulated events into   histograms with finite bins   which, if a smooth approximation 
to the target distribution is desired, can be  fitted with a parametric model.
An interesting question is whether it is possible 
to construct the smooth approximation directly 
from a Monte-Carlo simulation, bypassing 
histograms as  
an intermediate step.

One possible approach, that we investigate in this paper,   is to decompose the unknown target distribution into a complete orthonormal set of basis functions, and compute their  coefficients using conventional  Monte-Carlo methods.
These coefficients can be thought of as \emph{moments} of distributions, since they are given by their weighted integrals.
Importantly, this procedure is as efficient as constructing a histogram, and can be used to compute a large number of observables simultaneously. 

We note that this idea may  not be entirely new \cite{1986desd}. For example,  it   has been used in Ref.~\cite{10.1063/1.435057} to determine the distribution of scattering angles in molecular collisions, and 
it is possible that 
there are other examples that we are not aware of. 
At the same time, histograms  have become a  method of choice in theoretical high-energy physics  due to their  conceptual simplicity, robustness, and   ability to handle kinematic distributions spanning multiple orders of magnitude.  

A complete  set of basis  functions is, in general,  infinite-dimensional, and  in any practical calculation it needs to be truncated.
Although one may think that this is a challenge for accurately reproducing 
the target functions, we note that 
many kinematic distributions in high-energy physics are smooth and, roughly, have the shape of an asymmetric bell curve (transverse momenta distributions) or a smoothed rectangle (rapidity distributions).  Because of this, one can expect that a limited number of basis functions would be sufficient  to adequately describe them.

The method of moments has an additional key benefit when used with perturbative calculations based on local subtraction schemes \cite{Frixione:1995ms,Catani:1996vz,Gehrmann-DeRidder:2005btv,Czakon:2010td,Cacciari:2015jma,DelDuca:2016ily,Caola:2017dug,Magnea:2018hab,delduca2026nnlocalcompletelylocalsubtractions}.
Real and virtual corrections to differential cross sections exhibit  infrared singularities, when 
taken separately. 
Local subtraction schemes add terms to the real corrections to subtract these singularities in the soft and  collinear limits. From a computational Monte Carlo perspective,  subtraction terms
provide counter-events for each generated event,  making their  
combination integrable.  Although  these counter-events are only needed  in kinematic regions where additional radiation becomes indistinguishable from the no-radiation case, the counter-events  often extend to the whole phase space.  
Since the kinematics of an event and counter-events  differ, they may end up in separate histogram bins. 
Although this rarely happens, such events are, in some sense, 
catastrophic, and it takes a significant amount of  CPU time to recover from them.  

Moments, on the other hand, are largely immune to this issue. Indeed, since  weights that are used to calculate moments are continuous functions of the kinematic observables, all events and counter-events contribute to each moment, which provides a natural way to smooth the potentially offending contributions. 

An important question when using moments is the choice of basis 
functions. Obviously, there is significant freedom in their choice, which  can be exploited  to construct the most suitable basis for describing a particular target distribution. 
Indeed,  one can employ  known sets of orthogonal functions such as harmonic functions, Legendre or Chebyshev  polynomials etc. \cite{Szego1939Orthogonal,boyd01:CFS}. However, as we will argue below, this is often not \emph{an optimal choice}.  
Partially, this is related to the fact that  kinematic distributions   in collider physics
can change by several orders of magnitude on  the interval of interest. This feature is not conducive  to the reconstruction of  the target distribution from a relatively small number of moments. 

However, if a fairly  high-quality 
approximation to a particular  target distribution is available, 
it can be used  to construct an orthonormal basis that is optimized for describing this distribution  using moments.\footnote{We note that a similar idea was discussed in  Ref.~\cite{Ligeti:2008ac} in the context of designing a flexible
parametrization for the $B$-meson shape function. 
However, we believe that it was never applied 
for computing  higher-order QCD effects  in  
kinematic distributions relevant for collider physics.}
In the context 
of perturbative computations for collider physics, such 
approximations can be obtained from  
the leading-order distributions, which are  much easier to calculate than the same distributions at higher perturbative orders. 
From the perspective of moments calculations, leading-order results provide  high-quality approximations because 
QCD corrections change their shapes by tens of percent at most, 
whereas distributions 
themselves can change 
by orders of magnitude.
 
When such an  optimized basis is used, the first basis function is  proportional to the chosen initial approximation,
and other basis functions are variations needed to describe higher-order corrections to the shape of such a distribution.
In comparison to e.g.  Legendre moments, an optimized basis converges faster and avoids large fluctuations in regions 
where the target distribution is small. 

The paper is organized as follows. 
In  Section~\ref{sec:legendre_moments}
we introduce  the idea of moments as 
an alternative to histograms,  and provide the relevant mathematical details.
We  also discuss how to choose the highest moment to  
retain in the calculation, and estimate the truncation error based on the decay rate of the moments coefficients.
In Section~\ref{sec:toy} we compare how the method of Legendre moments\footnote{We have used different orthogonal polynomials and found 
rather similar results. For this reason, we will 
focus on Legendre polynomials in the rest  of this 
paper.} performs in comparison to conventional histograms for a number of toy Monte-Carlo examples.  In Section~\ref{sec:results} we apply these ideas to a   real-world example --  the calculation of the next-to-next-to-leading-order (NNLO) QCD corrections 
to Higgs boson production in weak-boson fusion (WBF). In Section~\ref{sec:optimized_basis} we show how to use an initial approximation to  the target distribution to construct an orthonormal basis that is optimized for computing it, and we apply this idea  to WBF. We conclude in Section~\ref{sec:conclusions}.

\section{Legendre moments}\label{sec:legendre_moments}

\subsection{Approximation of distributions with moments}

In this section we explain how to calculate the Legendre moments and use them to reconstruct a 
target distribution.
Suppose that we want to approximate an unknown target distribution $f(x)$ for 
a variable $x$ on an interval $x\in [a,b]$. We assume that for $x\in [a,b]$, there exists a complete set 
of functions $e_k(x)$, which are orthonormal 
\be
    \langle e_k,e_l\rangle=\delta_{kl},
\label{eq1}
\ee
with respect to a particular scalar product
\be
    \langle f,g\rangle=\int 
    \limits_{a}^{b} w(x) \; f(x) \; g(x) \; \mathrm{d}x.
    \label{eq2}
\ee
In Eq.~(\ref{eq2}) the  weight function $w(x)$  is 
arbitrary except for the fact that  it should be positive-definite, $w(x) > 0$,  on the interval $x\in [a,b]$. As we will explain below, the choice of $w(x)$ should be 
driven by  the features of the target distribution that we try to approximate.  Making a particular choice for $w(x)$  may  lead to a faster convergence of the moments-based approximation, especially in  regions where $w(x)$ is large.

Equipped with the scalar product Eq.~(\ref{eq1}), we can express the target distribution as follows
\be
    \label{eq:basis_expansion}
    f(x)=\sum_{k=0}^\infty
    c_k \; e_k(x),
\ee
where 
\be
    c_k = \left\langle
    f,e_k\right\rangle=\int\limits_a^b f(x)   \; w(x) \; e_k(x) \; \mathrm{d}x,
\ee
is the $k$-th  moment of the function $f(x)$ with respect to the chosen basis.

As we mentioned in the Introduction, we will 
consider Legendre polynomials as a prototypical example of orthogonal polynomials. 
They are defined as%
\footnote{Legendre polynomials can be efficiently evaluated using the Bonnet's recursion formula, and the Legendre series can be accurately evaluated using the Clenshaw algorithm~\cite{clenshaw1955note}.}
\be
    P_k(x)=\frac{1}{2^k k!} \; 
    \left ( \frac{{\rm d}}{{\rm d}x}
    \right )^k(x^2-1)^k.
\ee
Legendre polynomials  
form a complete basis 
on the interval $x \in [-1,1]$ and are orthogonal to each other
\be
    \int\limits_{-1}^1P_k(x)P_l(x)\mathrm{d}x=\frac{2 \delta_{kl}}{2k+1}.
\ee
To remap the Legendre polynomials to the interval $[a,b]$, we use a smooth monotonic variable transformation 
\be
t = t(x),\;\;\;
t(b) = 1, \;\;\; t(a) = -1.
\ee
Accounting for the normalization change, we choose the basis functions as 
\be
\label{eq:basis_functions}
    e_k(x)=\sqrt{k+1/2} \; t'(x) P_k(t(x)),
\ee
where we used the notation $t'(x) = {\rm d}t/{\rm d}x$. The functions in Eq.~(\ref{eq:basis_functions}) are orthonormal with respect to the weight function $w(x)=1/t'(x)$ on the interval $x  \in [a,b]$.
We note that 
the simplest choice for  $t(x)$ is the 
linear mapping 
\be
    t(x)=\frac{2x-a-b}{b-a},
\ee
in which case the weight evaluates to $w(x)=(b-a)/2$.

We now discuss the 
computation of  such moments for observables  in scattering processes 
at colliders. 
Our starting point is an observation that for a given set of basis functions $e_i$, the relevant moments can be  
evaluated using the standard Monte-Carlo methods.
Indeed, suppose that $\Phi$ describes the phase space of particles in a  scattering process. The observable ${\cal O}$ is a function of a phase-space point  $\phi \in \Phi$. 
The distribution of the variable ${\cal O}$ is given by 
\be
    \frac{\mathrm{d}\sigma}{\mathrm{d} {\cal O}} \Bigg |_{{\cal O} = x} =\int\limits_\Phi  \; \delta\left(x-{\cal O}(\phi)\right)
    \frac{\mathrm{d}\sigma}{\mathrm{d}\phi}\mathrm{d}\phi ,
\ee
where ${\rm d} \sigma/{\rm d} \phi$ is the 
differential cross section. 
Hence, the $k$-th moment 
of the ${\cal O}$-distribution in the interval ${\cal O}  \in [a,b]$ reads 
\be
    \label{eq:moments_phi}
    \left\langle\frac{\mathrm{d}\sigma}{\mathrm{d}{\cal O}},e_k\right\rangle
    =\int\limits_\Phi \; 
    \hspace{-0.2cm} \Theta^b_{a}({\cal O}(\phi) ) 
    w\left({\cal O}(\phi)\right)
    e_k\left({\cal O}(\phi)\right)
    \frac{\mathrm{d}\sigma}{\mathrm{d}\phi}\mathrm{d}\phi,
\ee
where 
\be
    \Theta^{b}_a(x)=\begin{cases}
        1&a<x<b\\
        0&\text{otherwise}.
    \end{cases}
    \ee

Integration over  $\Phi$ can be performed using the standard Monte-Carlo methods. In general, phase-space points $\phi \in \Phi$ are sampled from a probability density $ p(\phi)$. Therefore,  the integral over  $\Phi$ 
\be
   I =  \int\limits_\Phi f(\phi) \; \mathrm{d}\phi,
\ee
can be interpreted as the  expectation value of the function $f(\phi)/p(\phi)$, i.e. 
   \be
I = \mathbb{E}\left[\frac{f(\phi)}{p(\phi)}\right].
\ee
This  expectation value can be estimated by drawing $N$ random  samples of the phase-space points 
$\phi_i$, $i \in \{1,..,N\}$ from the probability distribution $p(\phi)$, and computing the average value 
\be
 \bar I 
= \frac{1}{N}
\sum \limits_{i=1}^{N} I_i,
\ee
where 
\be
I_i = \frac{f(\phi_i)}{p(\phi_i)}.
\ee

Then, the  integral $I$ is estimated using the following formula 
\be
 I = \bar I \pm \delta I,
\ee
where 
\be
\label{eq:MCuncertainty}
\delta I = \sqrt{
        \frac{1}{N-1}\left(
            \frac{1}{N}\sum_{i=1}^{N}I_i^2
            -\bar{I}^2
        \right)
    },
\ee
is the standard deviation. 

We note that many   moments 
for  the multitude  of 
observables 
can be evaluated in a \emph{single}  Monte-Carlo integration, using the same set of generated phase-space samples. 
In fact, the construction of a histogram can be viewed as an evaluation of the integral in Eq.~\eqref{eq:moments_phi} since 
a histogram with the bin extents $[x_{k,\text{low}},x_{k,\text{high}}]$ corresponds to rectangular functions
\be
    e_k(x)=
    \frac{\Theta^{x_{k,\text{high}}}_{x_{k,\text{low}}}(x)}{\sqrt{x_{k,\text{high}}-x_{k,\text{low}}}},
\ee 
with $w(x)=1$.
This set of functions is obviously  orthonormal but not complete.

\subsection{Truncation of moments }\label{sec:truncation}
In  any calculation that involves 
numerical computation of moments, the series in Eq.~\eqref{eq:basis_expansion} must  be truncated. 
This truncation causes an error in the reconstructed 
distribution that needs to be estimated.  In addition, 
the numerical calculation of moments results in 
uncertainties in their values, which also lead
to errors in the reconstruction of the target distribution. 

The two sources of uncertainties have an interesting 
interplay, which makes  
the number of terms that should be retained in the series 
in Eq.~(\ref{eq:basis_expansion}),  dependent  on the Monte-Carlo integration uncertainty.
This interplay happens because for any sufficiently smooth function $f(x)$ its moments $\langle f,e_k\rangle$ \emph{decrease} when $k$ increases.  At the same time,  the Monte-Carlo integration errors in the estimates of $\langle f,e_k\rangle$ do not decrease, and at some point the error becomes larger than the true value of the moment  $\langle f,e_k\rangle$.  It is then reasonable to expect 
that the optimal number of moments to include 
is determined by the moment whose ``natural''  value 
is of the order of the Monte Carlo integration error.
The clarification  ``natural'' here is important because it allows us to ignore  cases where the actual value of a particular  moment 
is accidentally close to zero, 
but higher moments are not small. 

To this end, we need to design a procedure to find 
``natural'' values of moments for a particular distribution.  
Our starting point is an observation that  Legendre moments of analytic  functions decrease exponentially~\cite{wang2012convergence}.
Therefore, we expect that the $k$-th  moment of 
a typical distribution can be estimated as 
follows
\be
    \left|\left\langle f,e_k\right\rangle\right|\sim c \; r^k,  
\ee 
where $c$ and $r<1$ are distribution-dependent constants.  

\begin{figure}
    \centering
    \includegraphics[width=\linewidth]{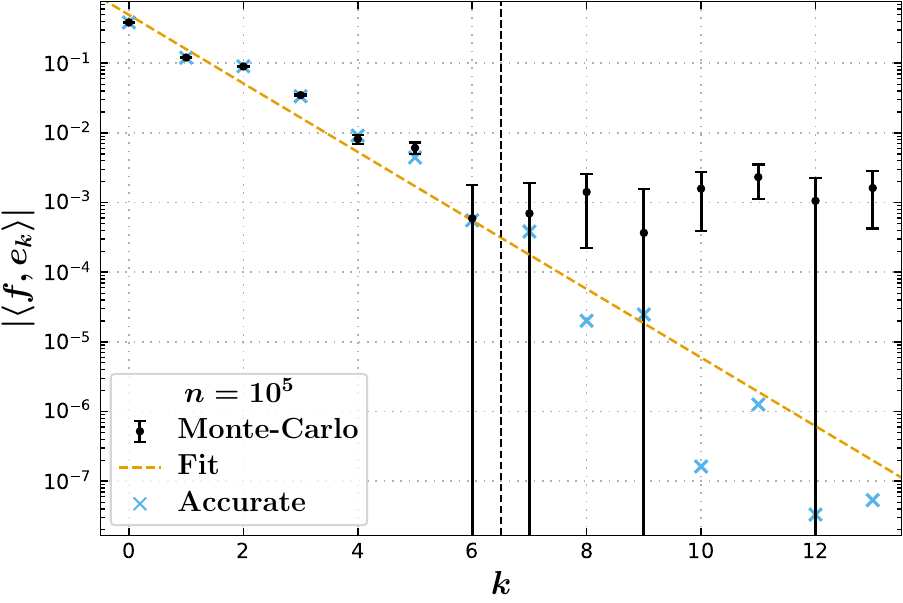}
    \caption{Legendre moments $\langle f,e_k\rangle$ of the normal distribution $f= 
    \mathcal{N}(0,1)$ (see Eq.~(\ref{eq22}) for the 
    definition)
    on the interval $[-1,2]$.
    The blue ``Accurate'' crosses are the true Legendre moments, calculated with high precision.
    The black data points are the moments calculated in a toy Monte-Carlo with $n=10^5$ samples.
    The orange dashed line is the estimate $\langle f,e_k\rangle\sim cr^k$ of the typical magnitude of the Legendre moments, obtained from the Monte-Carlo data (black points).
    The vertical gray dashed line marks the truncation point, determined according to the prescription described in Sec.~\ref{sec:truncation}.
    The distribution itself and its reconstruction from the retained moments are shown in Figure~\ref{fig:toy}.
    }
    \label{fig:moments_fit}
\end{figure}
We note that the sign of a moment  is  unknown a priori, and that  individual moments may, in fact,  be  much smaller than what this estimate suggests.  With this in mind, we model  moments  as independent normally-distributed random numbers with zero mean and variance $\sigma^2 = ( c r^{k})^2$.
Hence, we write 
\be
    \left\langle f,e_k\right\rangle\sim\mathcal{N}(0,cr^k),
\ee
where 
\be
{\cal N}(x_0,\sigma) 
= \frac{1}{\sqrt{2 \pi \sigma^2}} e^{-\frac{(x-x_0)^2}{2 \sigma^2}}.
\label{eq22}
\ee
The Monte-Carlo simulation yields moments $\hat\mu_k=\left\langle f,e_k\right\rangle+\varepsilon_k$ with some integration error $\varepsilon_k$ and an estimated variance $\delta\hat\mu_k^2$ of an  integration error. Then 
\be
    \varepsilon_k\sim\mathcal{N}(0,\delta\hat\mu_k),
\ee
for all moments $k$ with 
$ 0 \le k \le m-1$
for some $m$.
Hence, we assume that calculated moments are drawn randomly from the distribution 
\be
    \hat\mu_k=\left\langle f,e_k\right\rangle+\varepsilon_k\sim\mathcal{N}\left(0,\sqrt{c^2r^{2k}+\delta\hat\mu_k^2}\right),
\ee
where the coefficients $c, r$  are unknown. 
We then find them  by fitting the values of computed moments  $\hat\mu_k$. Technically, the fit is performed by numerically minimizing the log-likelihood function
\be
    \mathcal{L}(c,r)=\sum_{k=0}^{m-1}\left[\ln\left(c^2r^{2k}+\delta\hat\mu_k^2\right)+\frac{\hat\mu_k^2}{c^2r^{2k}+\delta\hat\mu_k^2}\right].
    \label{eq:likelihood}
\ee
This procedure yields an estimate of the ``natural'' size of moments $\left\langle f,e_k\right\rangle\sim cr^k$ that we use to compare against  the Monte-Carlo integration error. 
Finally, to determine the number  of moments 
that is retained for computing  the target distribution,  we find the  first moment $n+1$ for which  the inequality $\delta\hat\mu_{n+1}^2 > c^2r^{2(n+1)}$ holds. 
Then, Eq.~(\ref{eq:basis_expansion}) is computed 
by including   $n$ moments  in the sum on the right-hand side.  We note that 
the number $m$ of moments  that are calculated numerically by means of the  Monte-Carlo integration is chosen to be larger than the number $n$ of moments that are used to evaluate the target distribution.

We illustrate the use of this prescription with an example   in Figure~\ref{fig:moments_fit}, where we use moments to  
describe the function $e^{-x^2/2}/\sqrt{2 \pi} $ 
on the interval $x 
\in [-1,2]$.
Because the true values of moments $\langle f,e_k\rangle$ decrease with increasing $k$ while the Monte-Carlo errors tend to a constant,  there exists a 
moment $k^*$ whose Monte-Carlo estimate  
gets dominated by statistical noise.
In the current example this  happens around $k^* \sim 6$.

Monte-Carlo estimates of moments can  be considered reliable only to the left of this point, while to the right of it, the Monte-Carlo estimates are many orders of magnitude larger  than the true values.
The orange dashed line shows the trend $\langle f,e_k\rangle\sim cr^k$, with $c$ and $r$ estimated by minimizing the log-likelihood in Eq.~\eqref{eq:likelihood}. We remind the reader that this estimate is  based on the results of the Monte-Carlo computation  of the moments, but without the knowledge of the true values of $\langle f,e_k\rangle$.
In spite of this, the determined trend $\langle f,e_k\rangle\sim cr^k$ provides a reasonable, albeit slightly conservative  estimate of the possible  magnitudes of the true values of $\langle f,e_k\rangle$.
The vertical dashed line marks the truncation point determined by the prescription described above, --- in this case, the first $7$ moments are retained to provide an estimate of the target distribution.
We will revisit this example once again  in the next section.

Once the number of retained moments is  fixed 
and their asymptotic form is established, it becomes 
possible to estimate the total error of the 
approximation. To this end, 
for $x \in [a,b]$ we  estimate the basis 
functions as  $e_k(x) 
\sim \sqrt{1/(b-a)}$,  and compute  the remaining truncation error. We find 
\be
    \sum_{k=n}^\infty\left\langle f,e_k\right\rangle e_k
    \sim\sqrt{\sum_{k=n}^\infty\frac{c^2r^{2k}}{b-a}}
    =\frac{cr^n}{\sqrt{(b-a)(1-r^2)}}\,.
\ee
Then, the total approximation error is estimated as 
follows 
\begin{align}
    \label{eq:uncertainty}
    &\left|\sum_{k=0}^{n-1}\hat\mu_k e_k(x)-f(x)\right|
    \nonumber\\&\quad=\left|\sum_{k=0}^{n-1}\varepsilon_k e_k(x)
    -\sum_{k=n}^\infty\left\langle f,e_k\right\rangle e_k(x)\right|
    \nonumber\\&\quad
    \lesssim\sqrt{
        \sum_{k=0}^{n-1}
        \delta\hat\mu_k^2\left(e_k(x)\right)^2
        +\frac{c^2r^{2n}}{(b-a)(1-r^2)}
    }\,.
\end{align}
In the last step  we assumed that the Monte-Carlo errors in different coefficients are uncorrelated. Although this  is not exactly true, we find  it to be   acceptable to neglect the correlations between the moments for estimating the uncertainty. 

Higher basis functions tend to be more oscillatory, and the truncation procedure described above amounts to discarding higher frequencies in the approximate distribution.
This means that this truncation procedure is biased towards smoother distributions, which we actually see as a desirable form of regularization.

\section{Toy examples}\label{sec:toy}
To illustrate the main features of the method of  moments,
introduced in the previous sections, 
we elaborate further on  the  toy model  introduced 
there.  To this end, we  consider  the  normal distribution $\mathcal{N}(0,1)$ as the target distribution. We  attempt  to approximate it on the interval $x \in [-1,2]$. We  do this by sampling points uniformly and then using these samples, 
weighted by the value of the function $\mathcal{N}(0,1)$, to 
either construct a histogram-based  or 
a moments-based approximation to it.

\begin{figure*}[t]
    \centering
    \includegraphics[width=.49\linewidth]{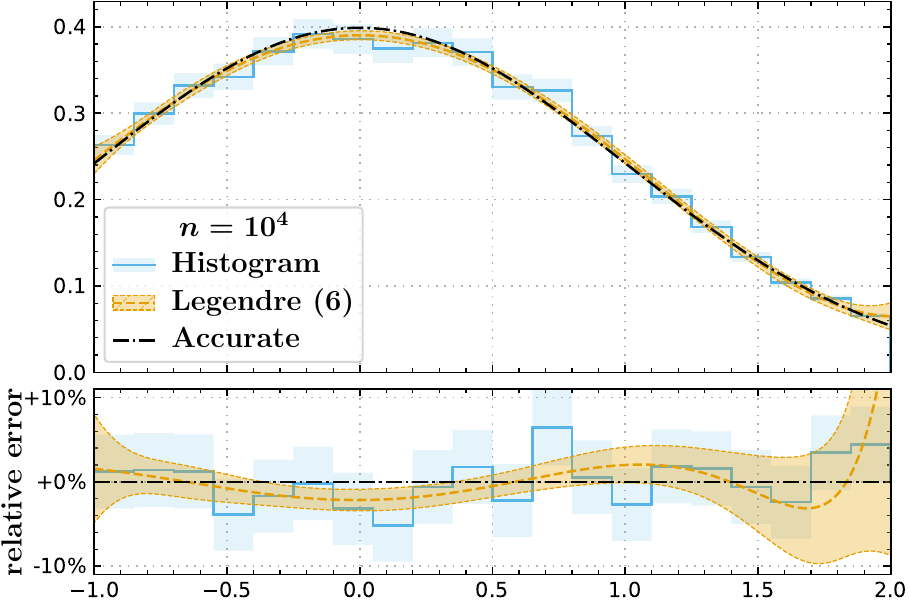}
    \includegraphics[width=.49\linewidth]{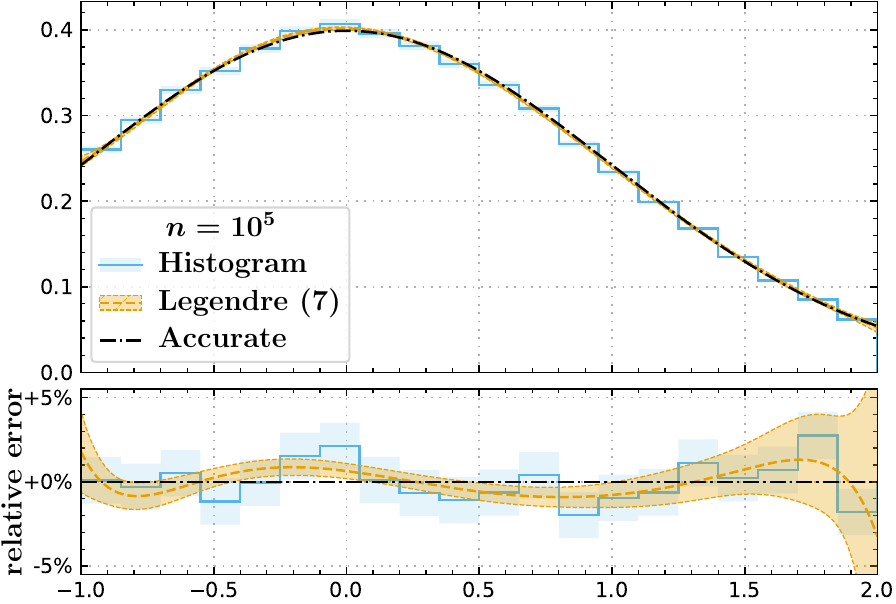}
    \\[3mm]
    \includegraphics[width=.49\linewidth]{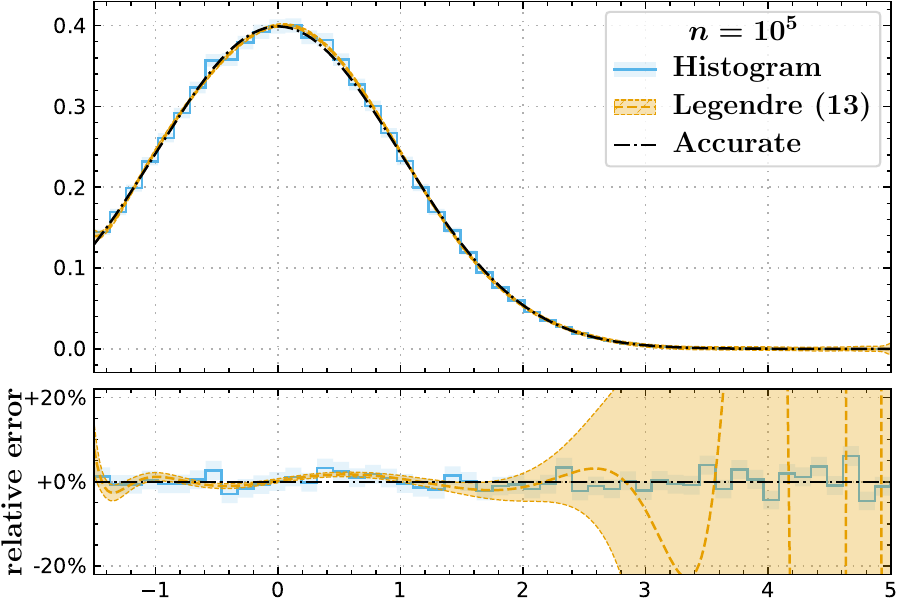}
    \includegraphics[width=.49\linewidth]{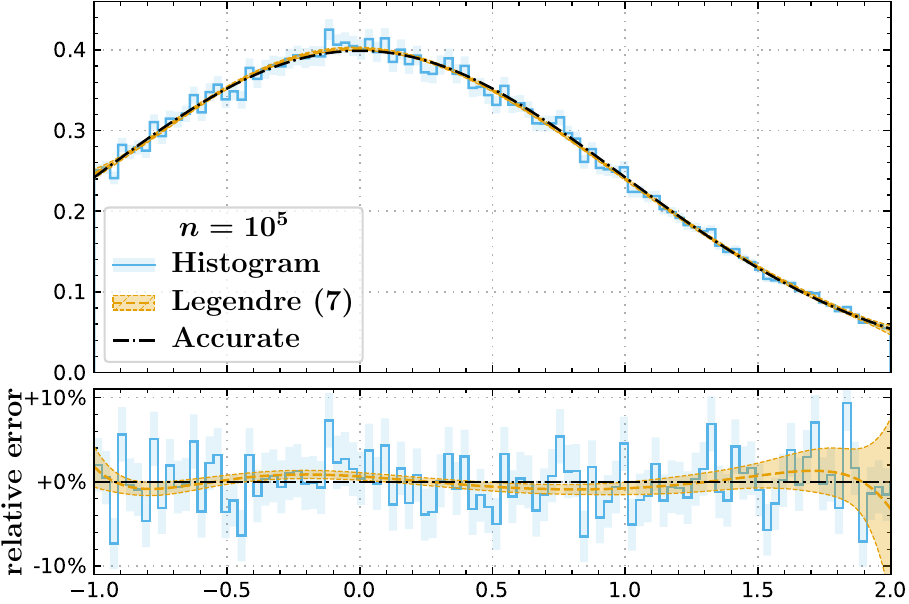}
    \caption{
    Normal distribution $(2\pi)^{-1/2}\exp(-x^2/2)$ reconstructed from $n$ samples using a histogram or Legendre moments.
    All plots, except 
    for the bottom-left one, compare the two approximations on  the interval $[-1,2]$.
    The bottom-left plot shows the approximation on the interval $[-1.5,5]$.
    The plots in the top row differ in the number of Monte-Carlo samples $n$.
    The histograms in the right column differ in the bin width, while the Legendre approximation is the same.%
    \commonCaptionA
    The bottom panels show the relative deviation of the reconstructed distributions from the true distribution.
    For each plot, the histogram and Legendre approximation are reconstructed using the same   Monte-Carlo samples.}
    \label{fig:toy}
\end{figure*}
Figure~\ref{fig:toy} shows the results for $n=10^4$ and $n=10^5$ Monte-Carlo samples.
The bottom panel in each of the four plots shows the relative deviation of the two kinds of approximations from the true result.
The number between parentheses in the legend for the moments is the number of retained moments after applying the truncation procedure described in Section \ref{sec:legendre_moments}.

The bands around the histograms show the Monte-Carlo integration uncertainty estimated according to Eq.~\eqref{eq:MCuncertainty}.
The uncertainty in the moments-based approximations 
is computed using  Eq.~\eqref{eq:uncertainty}; as we explained in the previous section, 
it accounts for both the Monte-Carlo and the truncation uncertainties.
We emphasize that for each plot 
in Figure~\ref{fig:toy}, the histograms and the moments-based  approximations are constructed from the same set of generated events.

The following features of various plots in Figure~\ref{fig:toy} are worth commenting upon.
\begin{itemize}
\item   With increased statistics, 
both methods converge to the true distribution.

\item Higher statistics allows us to compute more moments  reliably, and to  incorporate  them into  the moments-based approximation, according to the  truncation prescription that balances the Monte-Carlo and truncation errors.

\item  The differences between the histograms, the 
moments-based approximation and the true function are comparable in size 
\emph{around  the peak of the distribution.}

\item An important advantage of moments over histograms is that they yield smooth approximations of the target distribution. This is especially obvious when we compare them to a histogram with \emph{narrow bins}.
Such a comparison is shown in the bottom-right plot of Figure~\ref{fig:toy}.
With narrow bins, the number of events in each bin is smaller, and the statistical fluctuations are larger.
Moments, on the other hand, include information from all the generated samples, and they effectively smooth out these statistical fluctuations.

\item For large values of $x$,  the relative error of the moments-based distribution becomes very large, as illustrated in the bottom-left plot in Figure~\ref{fig:toy}. 
This is related to the fact that the approximation using standard polynomials provides 
a nearly uniform approximation error across the whole interval. The exponential suppression of the target 
distribution at large $x$ can be achieved by a precise
point-by-point cancellation of infinitely many 
contributions to the sum in Eq.~(\ref{eq:basis_expansion}).  The truncation,  
that we are forced to do,  violates this requirement 
and leads to a significant deterioration in the 
quality of the approximation based on moments. 
With higher statistics, the region where this problem occurs moves to the right since it becomes  
possible to include more  moments, but it never 
disappears completely. 
On the contrary, the relative error of a histogram is uniform, because the absolute error is proportional to the average value of the function in each bin, and we sample  points uniformly.

\end{itemize}
\begin{figure*}[t]
    \centering
    \includegraphics[width=.49\linewidth]{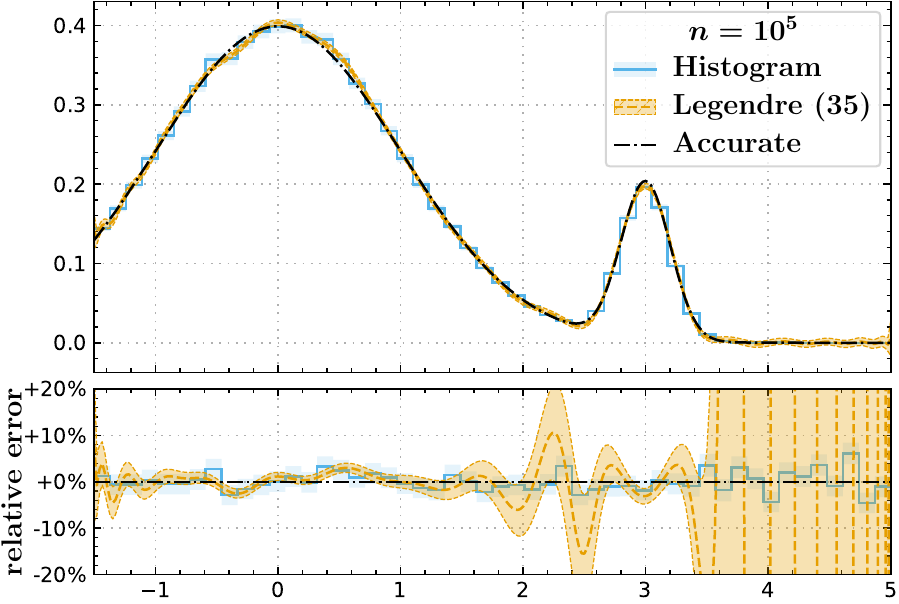}
    \includegraphics[width=.49\linewidth]{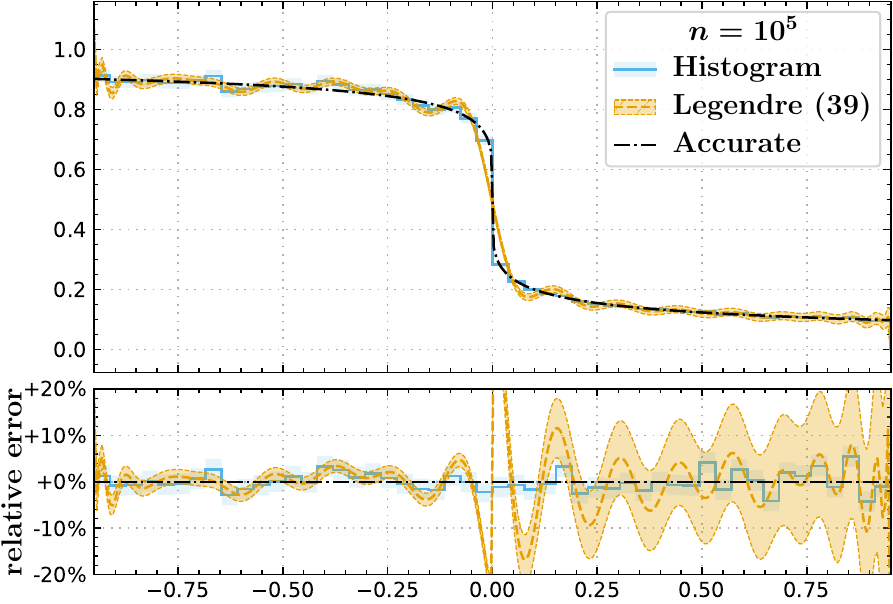}
    \caption{Examples of distributions that are more difficult to approximate using Legendre moments. The distribution on the left is a mixture of two gaussian peaks. The distribution on the right, defined in Eq.~\eqref{eq:smoothstep} has an infinite derivative at $x=0$.
    \commonCaptionA
    The bottom panels show the relative deviation of the reconstructed distributions from the true distribution.
    }
    \label{fig:toy2}
\end{figure*}
Throughout this discussion, we assumed that the target distribution $f(x)$ is smooth and does not possess any high-frequency features. If this assumption is violated, the moments-based approximation can be expected to 
perform worse. For illustration, consider the left panel in Figure~\ref{fig:toy2}, which displays a mixture of two Gaussians:
the first one  has mean zero, standard deviation one, and unit integral, whereas the second one  is centered at $x=3$, has a smaller  standard deviation of $0.2$, and the integral $0.5$. The presence of a second, relatively sharp peak makes it difficult to represent the overall distribution accurately using Legendre moments. In fact, we observe that the number of moments required for a reasonable description is several times larger than in the previous examples. Moreover, the relative approximation error remains elevated even in the regions far from the second peak.
\vskip1cm
Another pathological example is shown in the right plot of Figure~\ref{fig:toy2}. It refers to the following  target distribution
\be
    f(x)=\begin{cases}
        1-\frac{\alpha}{1+\sqrt{\beta|x|}}&x<0,\\
        \hphantom{1-{}}\frac{\alpha}{1+\sqrt{\beta|x|}}&x>0,
    \end{cases}
    \label{eq:smoothstep}
\ee
with parameters $\alpha=0.4$ and $\beta=10$.
This function has an infinite derivative at $x=0$.
Again, many Legendre moments are needed to describe this function, and the approximation error is relatively large.

A possible way to deal with such problems is to split the   $x$-interval into several sub-intervals, 
and construct the moments-based approximation 
for each sub-interval separately.  We did not pursue 
this issue further in the current study. Instead, 
we continue with the discussion of how the 
moments-based approximation fare in practice, and 
how the issue of large uncertainties in tails 
of distributions can be addressed.

\section{Moments in practice}\label{sec:results}

After the toy model, we consider a realistic example. As  mentioned in the introduction, our 
primary interest lies in  perturbative QCD calculations for collider processes, and the example 
that we would like to discuss falls into this category. 
We implemented the method of moments in the NNLO Monte-Carlo integrator for Higgs boson production in weak-boson fusion (WBF) that was developed in Ref.~\cite{Asteriadis:2021gpd}.
The input parameters and the WBF event selection criteria are the same as in Ref.~\cite{Asteriadis:2021gpd}.  We will mostly focus 
on the distributions of the transverse momentum $p_{\perp,H}$ and the rapidity $y_H$ of the 
Higgs boson, produced in the weak-boson-fusion, 
since these two distributions provide archetypal examples.

\begin{figure*}
    \centering
    \includegraphics[width=.49\linewidth]{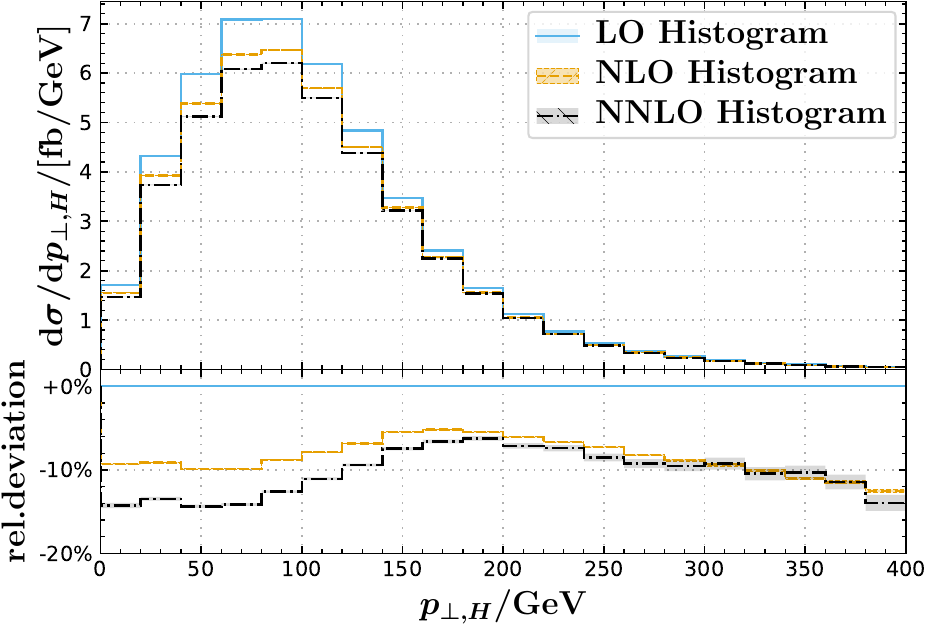}
    \includegraphics[width=.49\linewidth]{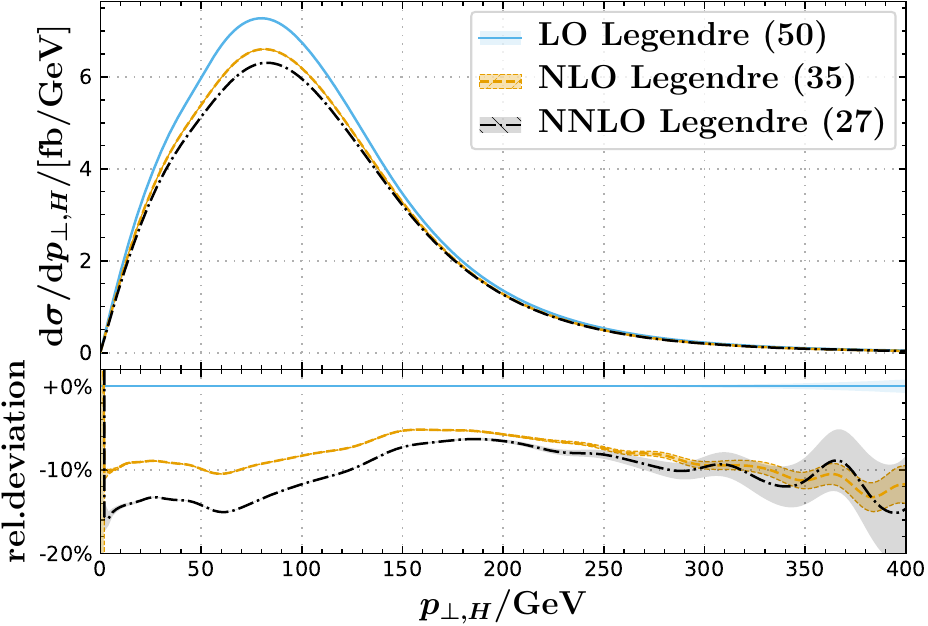}
    \caption{\commonCaptionB{transverse momentum $p_{\perp,H}$}}
    \label{fig:ptH}
\end{figure*}
\begin{figure*}
    \centering
    \includegraphics[width=.49\linewidth]{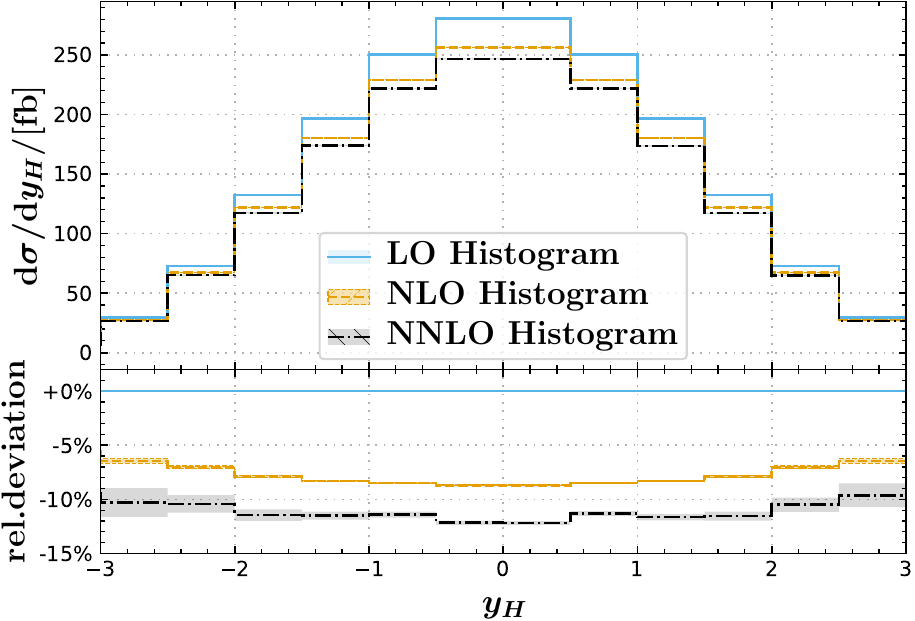}
    \includegraphics[width=.49\linewidth]{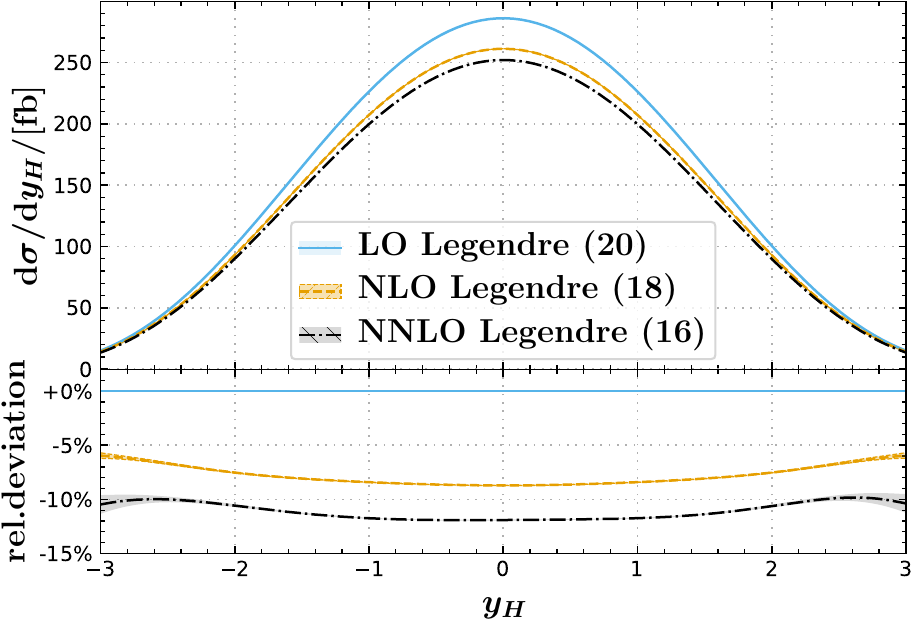}
    \caption{\commonCaptionB{rapidity $y_H$}}
    \label{fig:yH}
\end{figure*}
Figures~\ref{fig:ptH} and~\ref{fig:yH} show the $p_{\perp,H}$ and $y_H$ distributions at different perturbative orders, reconstructed using histograms and  Legendre moments.
The distributions, obtained with two different  methods, agree within the uncertainties.
As we have already discussed in the toy examples,  distributions reconstructed from the Legendre series have a larger relative error in the tails of the distributions.

We can use this example to illustrate one of the  key strengths of the method of moments, 
which is related to the fact that moments provide 
a natural way to smooth distributions. 
Indeed, as  we have mentioned previously, in calculations that employ  local subtractions,  contributions of events with additional real partons  need to be combined  with corresponding counter-events to ensure cancellation of infrared singularities. Since events and counter-events can have somewhat different kinematics, they can end up in different  histogram bins.
In this case the necessary cancellation between events and counter-events does not happen,  leading to  to a dramatic fluctuation of results recorded in neighboring bins.
\begin{figure*}[t]
    \centering
    \includegraphics[width=.49\linewidth]{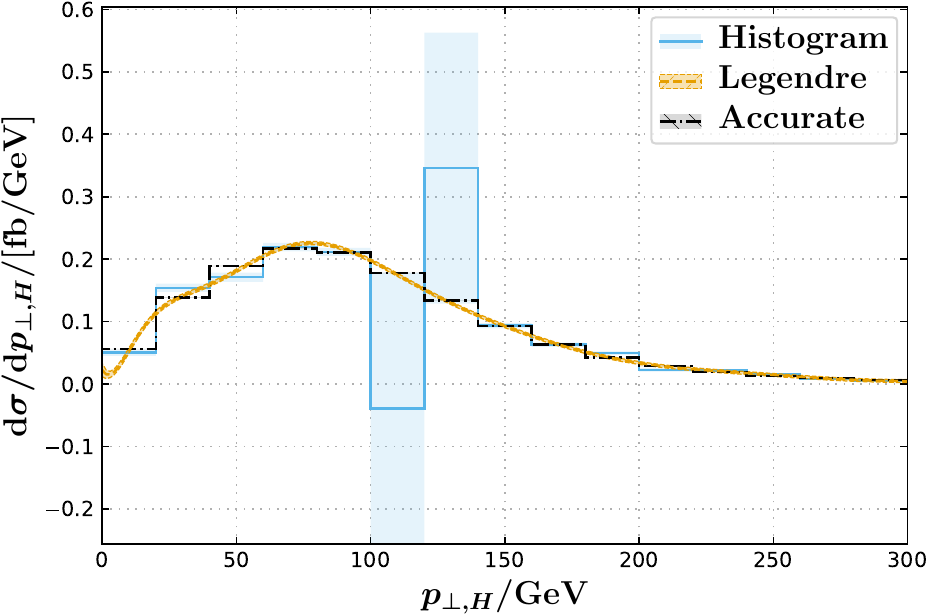}
    \includegraphics[width=.49\linewidth]{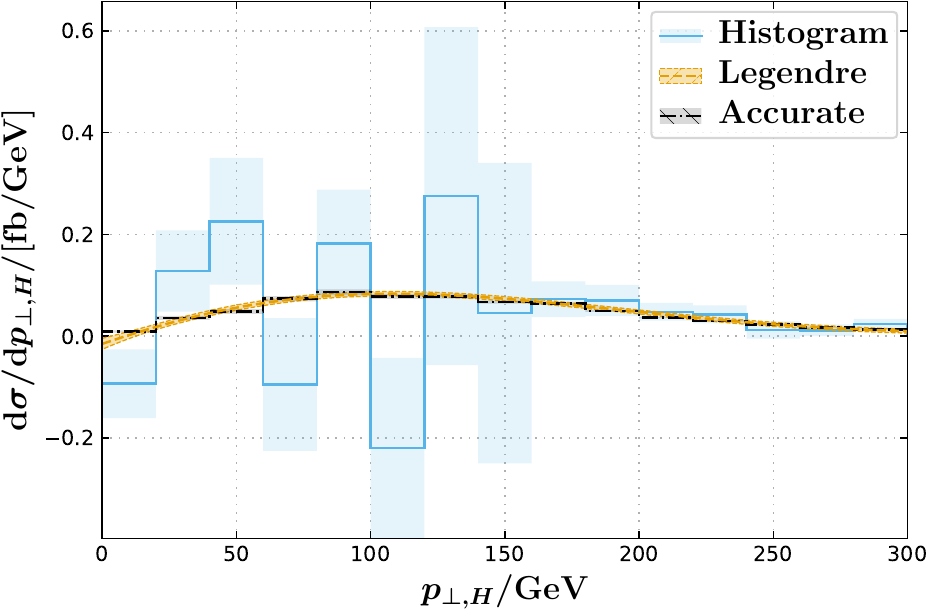}
    \caption{Examples of bin-to-bin fluctuations in some infrared-subtracted double-real NNLO corrections to WBF. The solid blue histogram (``Histogram'') and the orange dashed line (``Legendre'') show the correction reconstructed from the same set of generated events using a conventional histogram and using Legendre moments, respectively. The black dash-dotted histogram (``Accurate'') serves as a proxy for the true distribution, as it was generated using over 100 times more statistics than in the other two approximations. The left plot shows a prominent example of a bin-to-bin fluctuation, where an event and a counterevent landed on opposite sides of the bin edge at $p_{\perp,H}\approx\qty{120}{\GeV}$.
    The right plot shows a different infrared-subtracted double-real correction, which is dominated by such fluctuations.}
    \label{fig:fluctuations}
\end{figure*}
This phenomenon is illustrated in Figure~\ref{fig:fluctuations}, with examples taken from some distributions  contributing to the NNLO double-real corrections to WBF.

This problem is related to the fact that the integral over a bin of a histogram is a scalar product with a \emph{discontinuous} function,  the indicator function of the bin.
As a result, at the border between the  neighboring bins  soft and  collinear radiation can change to which bin the given event is assigned\footnote{The same issue exists at the boundaries of the acceptance regions needed for computing fiducial cross sections.}.
Although this may  sound like  a  violation of the infrared safety,  
since this problem only occurs  at  the edges of histogram bins, which form a  measure-zero  subset of the available phase space, this phenomenon does not render the integral divergent. However,  it does cause problems in numerical simulations with a finite number of samples since  once such a mis-cancellation  happens,  significant statistics is required  to overcome it, and converge to the right  result.

The difference between kinematics in the event and  the counter-event is related to the choice of the phase-space parameterization.
It may  be possible to design a phase-space parametrization that optimizes the choice of kinematics for counter-events, ensuring that the value of this observable is nearly  the same in the event and all corresponding counter-events.
However, even if one succeeds in finding a parametrization that  has this property for one observable, it is not guaranteed to hold for the other ones. 
Designing  a specialized phase-space parametrization for each considered observable  is impossible in practice. 

A  possible way to mitigate bin-to-bin fluctuations is to distribute an event weight across different bins, a procedure which is usually referred to as ``bin smearing''. Different variants of bin smearing have been proposed, including  fractional filling \cite{Bierlich:2019rhm} and joint manipulation of the weights of the events and counterevents using the understanding of the singularity structure of the integrand \cite{Sherpa:2024mfk}. Although these techniques are powerful, they introduce yet another smearing 
of original distributions, which the method of moments strives to avoid.

In fact, moments  are largely immune to mis-cancellations between events and counter-events. Because moments are scalar products with \emph{continuous} basis functions, a small change in the kinematic of the observable computed for  the event and the counter-event  leads to a similarly small change in their contributions to the moments.
Correspondingly, the distributions reconstructed from Legendre moments  do not show any sign of fluctuations between histogram bins, see  Figure~\ref{fig:fluctuations}.
We emphasize that the histograms and the Legendre series compared in each plot in Figure~\ref{fig:fluctuations} are constructed in the same  Monte-Carlo run using identical  samples.

\section{Basis optimization}\label{sec:optimized_basis}

The challenge in the moments-based 
reconstruction approach, that we  have 
witnessed  repeatedly   in the previous sections, is that their 
performance deteriorates in exponentially-suppressed tails of distributions. The goal of this 
section is to address this issue. In particular, 
 we will explain how a \emph{reasonably-accurate approximation to the target distribution}  can be used to construct an optimized basis for computing it.   In the  context of perturbative  
calculations, reasonably-accurate approximations 
to target distributions can often be inferred from 
the leading-order computations of the corresponding observables.

In Section~\ref{sec:legendre_moments} an orthonormal basis was constructed by mapping the target distribution from the interval $[a,b]$ to the interval $[-1,1]$ 
with the help of 
a linear variable transformation, and 
then using the Legendre polynomials   in the transformed coordinates as the basis functions. 
The main idea behind the basis optimization is to trade a linear variable transformation for a non-linear one, 
and to choose it in such a way that a reasonably-accurate approximation to the  
target distribution becomes flat in the new coordinates. 

Hence,  if we are interested in the kinematic 
distribution characterized by a variable $x$ defined 
on the interval $x \in [a,b]$,  and if $f_0(x)$ is the 
approximate target distribution,  we choose
\be
    t(x)=-1+2\frac{\int\limits_a^x f_0(\tilde{x})\mathrm{d}\tilde{x}}{\int\limits_a^bf_0(\tilde{x})\mathrm{d}\tilde{x}},
    \label{eq:rectifying_function}
\ee
as the variable transformation $x \to t$.
This transformation  can only be employed if  the function $f_0(x)$ is sign-definite since, otherwise,  the variable transformation $x \to t $ is not invertible.

We define the basis functions as follows  
\be
    e_k(x)=\sqrt{k+1/2}\; t'(x) P_k(t(x)).
    \label{eq30}
\ee
It is easy to see that these functions 
are orthonormal on the interval $x \in [a,b]$ with  the weight function
\be
    w(x)=\left(t'(x)\right)^{-1}
    \propto\left(f_0(x)\right)^{-1}.
\ee
The completeness of this basis follows immediately from the completeness of Legendre polynomials.
To distinguish the optimized basis from the standard Legendre polynomials, we call this basis ``rectified''.

Suppose that the kinematic distribution $f_0(x)$ differs 
from the true distribution $f(x)$ by a function $K(x)=f(x)/f_0(x)$. If the initial approximation $f_0(x)$ is the leading-order distribution, then $K$ plays the role of a differential $K$-factor, which for many perturbative computations  is a slowly-varying order-one function.\footnote{The function  $K(x)$ is  ``slowly-varying''  when  compared to possible exponential changes of target distributions $f(x)$ and 
$f_0(x)$. } The moments of $f(x)=K(x)f_0(x)$ then evaluate to  
\be
c_k = \sqrt{k+1} \;
\frac{\int \limits_{a}^{b} {\rm d} x f_0(x) }{2}
\int \limits_{-1}^{1} {\rm d} t P_k(t) 
K(x(t)).
\ee
Hence, if $K(x(t))$ is flat, only a relatively small 
number of Legendre polynomials will significantly differ from zero. 

\begin{figure}
    \centering
    \includegraphics[width=\linewidth]{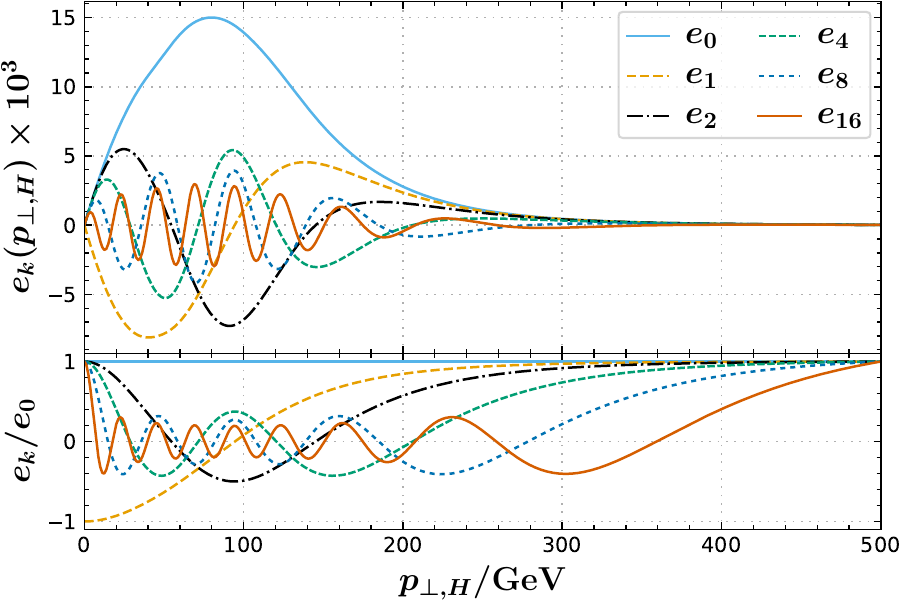}
    \caption{Basis functions constructed from an initial approximation of the distribution of Higgs boson transverse momentum in weak-boson fusion. The bottom panel illustrates the relative difference between different basis functions.}
    \label{fig:ptH_basis_functions}
\end{figure}
To give an example, in Figure~\ref{fig:ptH_basis_functions} we show the basis functions $e_k$ for the transverse momentum distribution of the Higgs bosons produced in the weak-boson fusion. We consider the transverse momenta on the interval $0 < p_{\perp,H} < 500~{\rm GeV}$.  The function 
$f_0(x)$ is identified with the leading-order transverse momentum distribution. 
We approximate it by  a truncated sum of Legendre polynomials calculated as in the previous sections.
This can be easily done because leading-order distributions are typically inexpensive to calculate, which means that they  can be accurately described, even if they are not known analytically.

It follows from Eq.~(\ref{eq30}) that 
the first basis function $e_0(x)$ is proportional to 
$f_0(x)$, and all other basis functions feature the exponential
suppression of the high-$p_\perp$ region as well.
Although  basis functions do oscillate, 
as dictated by the presence of the Legendre polynomials 
in Eq.~(\ref{eq30}),  these oscillations  are mostly confined to the region around the peak of the distribution. Therefore, one can expect that this set of basis functions will  describe fine features of the target distribution around its  peak,  and provide a smooth approximation to the target distribution in its tail.

\begin{figure*}
    \centering
    \includegraphics[width=.49\linewidth]{plots/ptH_legendre.pdf}
    \includegraphics[width=.49\linewidth]{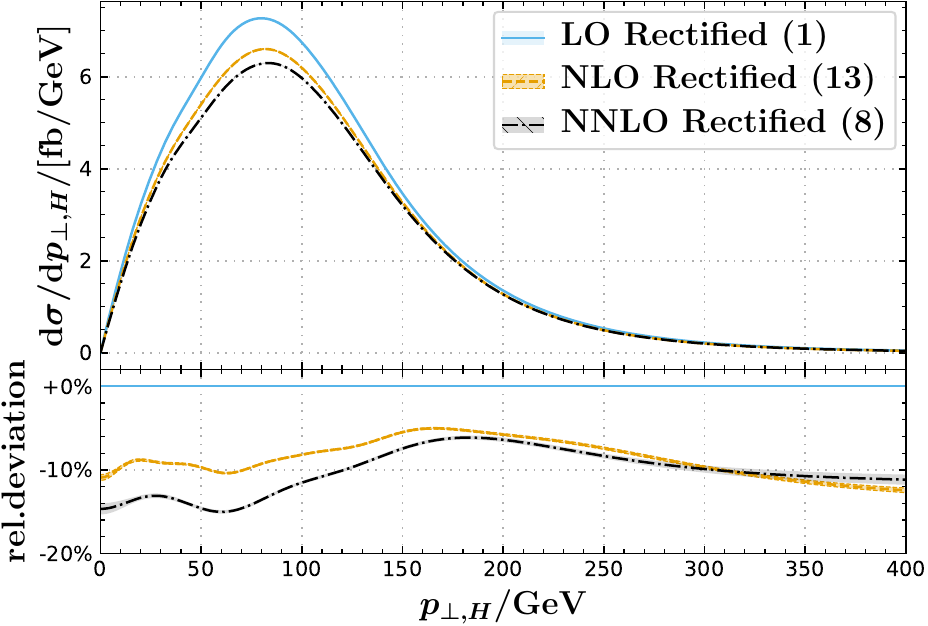}
    \caption{Distribution of transverse momentum of the Higgs boson approximated using using Legendre moments (left) and moments in a ``rectified'' basis constructed from the leading-order approximation (right), at different perturbative orders.
    \commonCaptionA
    The bottom panels show the relative deviation of the reconstructed distributions from the leading-order distribution. Significant improvement in the quality of the  description of the high-$p_{\perp}$ tail of the distribution is observed when the rectified basis is used.  
    }
    \label{fig:ptH_rectified}
\end{figure*}
A comparison of the calculation of the Higgs boson transverse momentum distribution  based on the standard basis of Legendre polynomials, and the calculation with the optimized basis, is presented  in Figure~\ref{fig:ptH_rectified}.
We note that the number of moments that  
are needed to adequately describe the distribution 
in the two cases differs by nearly a factor three, 
indicating faster convergence if the optimized basis is used. 
Because all functions in the optimized basis are constructed to reproduce the exponentially-suppressed tail of the leading-order result, the calculated 
distributions at higher orders become proportional 
to $f_0(x)$ in the tail and do not exhibit any large fluctuations.

It should be appreciated, however, that  this feature 
is a consequence of the way the optimized 
basis is constructed. This basis is not inherently more sensitive to the tail of the distribution; rather it effectively maps a 
significant  interval in Higgs $p_\perp$  to a narrow 
region in $t \in [-1,1]$ interval so that any features in the high-$p_\perp$ region  are averaged 
over.  Conceptually, this is equivalent to taking histograms with larger bins in regions where the 
target distribution is small, and this may lead to the distortion of the shape of the distribution at high $p_\perp$. 

\begin{figure*}
    \centering
    \includegraphics[width=.49\linewidth]{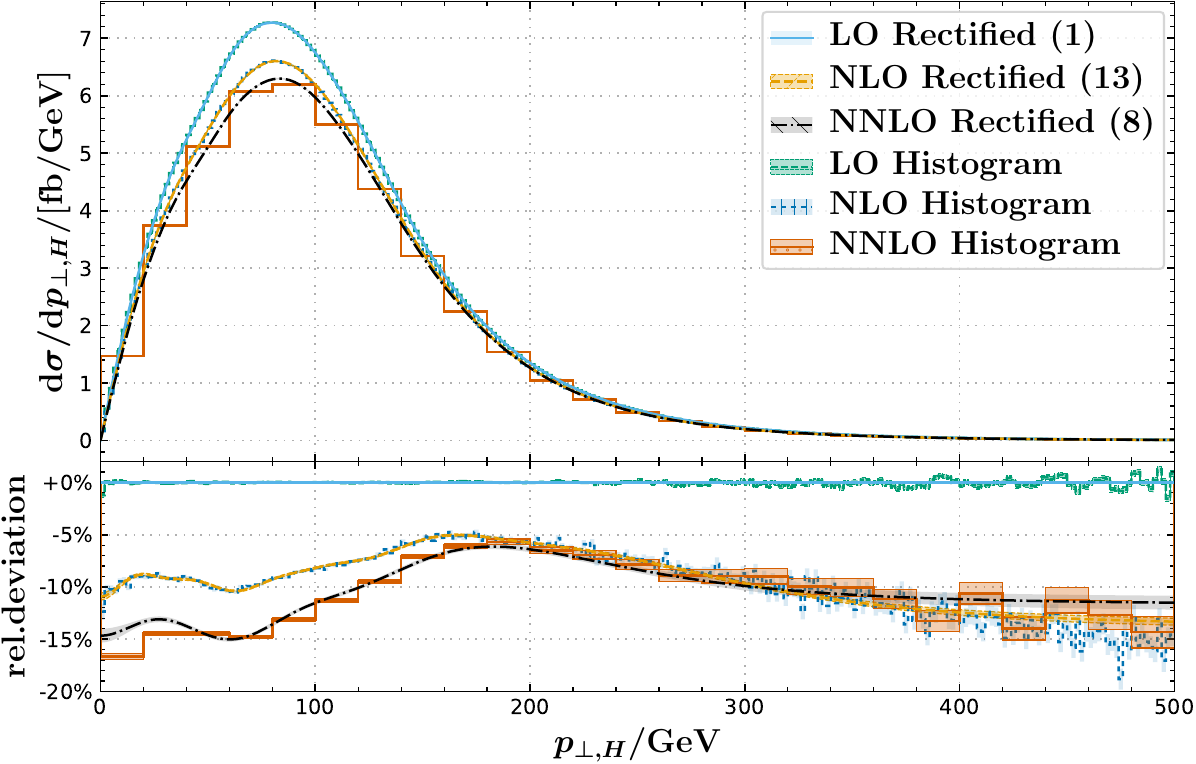}
    \includegraphics[width=.49\linewidth]{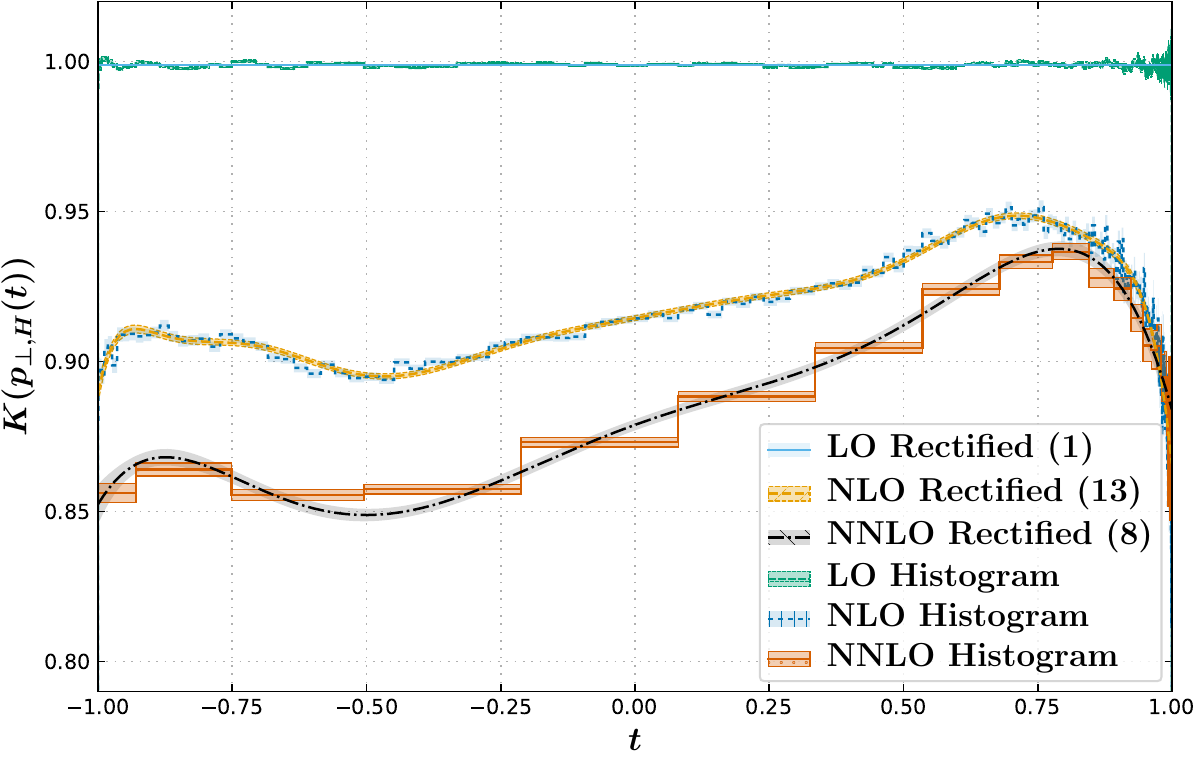}
    \caption{
    The left plots shows the distribution of transverse momentum of the Higgs boson in weak-boson fusion approximated using conventional histograms or moments in a ``rectified'' basis constructed from the leading-order approximation.
    The right plot shows the same distribution in the transformed variable $t(p_{\perp,H})$, constructed according to Eq.~\eqref{eq:rectifying_function}.
    The bin size for the histograms is \qty{2}{\GeV} at leading and next-to-leading orders and \qty{20}{\GeV} at NNLO.
    \commonCaptionA
    The bottom panel on the left plots shows the relative deviation from the leading-order distribution.}
    \label{fig:ptH_vs_fine_bins}
\end{figure*}
To illustrate this point, in the left plot of Figure~\ref{fig:ptH_vs_fine_bins} we compare the distribution reconstructed using the optimized basis against conventional histograms with very fine bins at leading  and next-to-leading order.
At NNLO we have to use coarser binning, because at this perturbative order bin-to-bin fluctuations become 
overwhelming if small bins are employed.
Around the peak of the distribution the moments approximations are very accurate and can resolve even very fine features of the shape modification due to higher-order corrections, but the resolution deteriorates towards the tail of the distribution.
The histograms and the moments approximations agree within uncertainties for $p_{\perp,H}\lesssim\qty{350}{\GeV}$, but above this point some discrepancy in the slope of the reconstructed distributions is visible at NLO. The smoothing of the true distribution that arises 
because of the use of moments appears to  be too aggressive in this region.

The origin of this problem can be seen in the right plot in Figure~\ref{fig:ptH_vs_fine_bins}, which shows the same distribution in the \emph{transformed coordinates}.
We note that in these coordinates
the optimized basis functions reduce to the standard Legendre polynomials. However, when  written  in  
$t$-variables, functions  that are relatively smooth  in the $p_\perp$-variable  start changing very rapidly   
near the endpoint $t=1$. This feature is not well-described by a few Legendre polynomials, which translates to an overly aggressive smoothing in the far-tail region of this observable. One way to solve this problem is to exclude the region 
where the distribution in the $t$-variable starts 
to decrease rapidly. In the current case, 
this corresponds to the right-most ${\cal O}(5\%)$ of the interval $t \in [-1,1]$, which translates to $p_{\perp,H} \gtrsim \qty{300}{\GeV}$. 

We note that a similar  issue is  also present in other kinematic distributions that we studied,  but it was not as prominent as in the example we show in Figure~\ref{fig:ptH_vs_fine_bins}.  We also stress that it is important that this problem can be easily detected by looking at a  particular distribution in transformed coordinates and a ``safe'' region in 
the original coordinates can be identified. 

Finally, we note that the approaches discussed above are not unique. A variety of alternative reconstruction methods exist, including different functional bases or target distribution manipulations. We explored several such possibilities and obtained results of varying quality. For brevity, and because these alternatives do not qualitatively change the conclusions of our work, we do not discuss them  
here.

\section{Conclusions}\label{sec:conclusions}

We have shown  that moments-based approximations provide an attractive  alternative to conventional histograms.
They directly yield smooth approximations to the target distributions, and, in perturbative computations based on  local subtraction schemes, they are immune to large fluctuations that arise when an event and a counter-event  
get  accommodated into neighboring bins.

For smooth distributions with no high-frequency features, moments-based approximations converge faster than conventional histograms of comparable resolution.
Furthermore, they yield compact representations of distributions that do not depend on the arbitrary choice of binning, admit efficient evaluation and integration, and are inherently differentiable, which might be useful for applications such as  
gradient-based optimization. Furthermore, 
it is straightforward to use moments to recover conventional histograms, for any desired binning.

Moreover, if an informed  high-quality approximation to the target distribution  is available, it can be used to construct an optimized basis of functions and moments.  
As we have shown, this leads to faster convergence and smooths  out fluctuations in the tails of the reconstructed distributions.

While in this study we focused on one-dimensional distributions, the discussed methods generalize straightforwardly to multidimensional ones. We  expect them to perform similarly well, as long as the dimension is not too high.

In conclusion, moments-based methods of reconstructing distributions from random samples can outperform conventional histograms when the underlying distributions are sufficiently smooth and when calculations are performed with local subtraction schemes. 
For these reasons, we advocate for further exploration and use of these methods in Monte-Carlo integrators.

\begin{acknowledgments}

We are grateful to K.~Asteriadis, A.~Behring and 
R.~R\"ontsch for useful conversations. This research is partially supported by the Deutsche Forschungsgemeinschaft (DFG, German Research Foundation) under grant 396021762 - TRR 257.
\end{acknowledgments}

\bibliography{moments}
\end{document}